# SPIN WAVES ALONG THE EDGE STATES

A. LARA

*Departamento de Física de la Materia Condensada, Universidad Autónoma de Madrid*
*28049 Madrid, Spain*
*antonio.lara@uam.es*

V. METLUSHKO

*Department of Electrical and Computer Engineering, University of Illinois at Chicago*
*Chicago, 60607 Illinois, USA*
*vmetlush@ece.uic.edu*

M. GARCÍA-HERNÁNDEZ

*Instituto de Ciencia de Materiales ICMM, CSIC, Ciudad Universitaria de Cantoblanco*
*28049 Madrid, Spain*
*marmar@icmm.csic.es*

F. G. ALIEV[*]

*Departamento de Física de la Materia Condensada, Universidad Autónoma de Madrid, Address*
*28049 Madrid, Spain*
*farkhad.aliev@uam.es*

Spin waves have been studied experimentally and by simulations in 1000 nm side equilateral triangular Permalloy dots in the Buckle state (B, with in-plane field along the triangle base) and the Y state (Y, with in-plane field perpendicular to the base). The excess of exchange energy at the triangles edges creates channels that allow effective spin wave propagation along the edges inthe B state. These quasi one-dimensional spin waves emitted by the vertex magnetic charges gradually transform from propagating to standing due to interference and(as pointed out by simulations) areweakly affected by smallvariations of the aspect ratio(from equilateral to isosceles dots) or by interdot dipolar interaction present in our dot arrays. Spin waves excited in the Y state have mainly a two-dimensional character.Propagation of the spin waves along the edge states in triangular dots opens possibilities for creation of new and versatile spintronic devices.

*Keywords*: Spin waves, edge states, magnetic dots.

[*] Corresponding author: farkhad.aliev@uam.es.

## 1. Introduction

Transport of intrinsic angular momentum of electrons spin over long distances is one of the challenges of magnonics, spin calorics and spintronics.[1-5] Using spin orbit (SO) effects in SO-metal/magnetic insulator/SO-metal structures spin wave (SW) transmission over macroscopic distances has been recently achieved and explained.[6,7] As to metallic structures, the possibility of exciting and manipulating SW has been usually linked with the implementation of low damping materials such as Heusler compounds (see Ref. 8) or $Fe_{1-x}V_x$ alloys (Ref. 9) and less attention has been paid to searching for new mechanisms for control over SW in magnetic materials already widely used, such as Permalloy. An effective alternative approach could be to explore control over the SW dimensionality through their confinement to the inhomogeneous edge states that exist in perpendicular to the edge in plain bias field in non-elliptical magnetic elements.[8-10] However, the edge states in rectangular magnetic elements (to our best knowledge the only field configuration where edge states have been investigated so far) are interrupted by the topological magnetic charge.[11] As a consequence, the edge SWs (E-SWs) are suppressed near the element center (Ref. 10) due to the above symmetry imposed interruption of the edge state.[11]

Here we study experimentally and via simulations static and dynamic (broadband response) magnetization in 1000nm side length triangular dots which, being larger than those investigated before (Refs. 12 and 13), have well defined edge and vertex states. Micromagnetic simulations identify the modes excited in the B state as SW emitted by the vertex magnetic charges that propagate and interfere along the edge states. Remarkably, the one dimensional (1D) character of these SW is qualitatively distinct from those observed previously for SWs in 1D magnonic crystals (Ref. 14), 2D SW diffraction (Refs. 15 and 16) or SW interference patterns from point contact spin torque emitters.[16-19]

Our finding experimentally confirm a number of recent theoretical predictions (Refs. 20-23) of the existence of a novel type quasi one dimensional SW propagating along the edges of ferromagnets, magnetic edge states in graphene and other magnetic structures with magnetic exchange energy stored near the device edges. Some initial results have been presented in Ref.24. Here, we provide further experimental and simulation details to support excitation of edge spin waves.

## 2. Experimental Details

### 2.1. *Sample preparation*

The samples studied consist of arrays of Permalloy triangular dots, covered with a gold coplanar waveguide (CPW) that generates a high frequency magnetic field (rf field). The dots were deposited on a Si (100) substrate by electron beam lithography and lift-off technique and they are equilateral triangles of 1000 nm side length and 30 nm thick, with an interdot separation of 200 nm in vertical direction (from top vertex of a triangle to base of the upper neighbor) and in horizontal direction (from bottom left vertex of a triangle to bottom right vertex of its neighbor), as shown in Fig. 1 a).

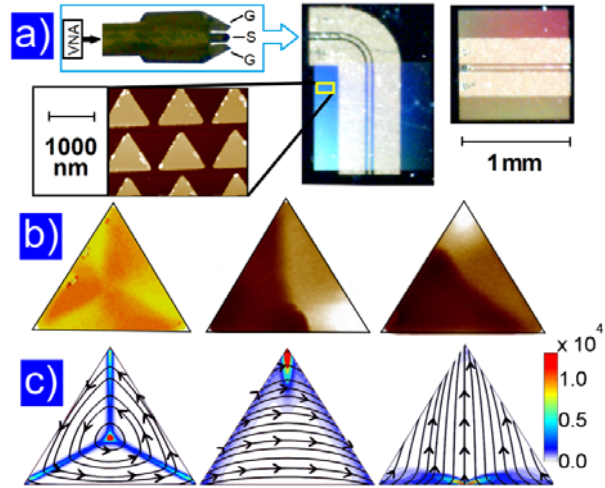

Fig. 1. a) Photograph of the high frequency probe that inputs the signal from the vector network analyzer (VNA) to the coplanar waveguide. "G" and "S" stand for "ground" and "signal", respectively. A picture of the two types of CPW is shown to the right, with and without a corner. The bright square under the CPW is the array of dots. A SEM image of the dots is attached.
b) MFM pictures at zero field. A vortex state (after creating the sample) is shown in the left, and the central and right pictures show the remanent state after saturating the dot parallel to the base (B) or perpendicular to the base (Y) and going back to zero field.
c) Micromagnetic simulations of V (at H=0 Oe), B (at H=1000 Oe) and Y (at H=1000 Oe) states. Arrows show the direction of magnetization, and the color scale represents the density of exchange energy (in $J/m^3$).

The CPW is 500 nm thick consists of a central electrode 30 μm thick, separated of two ground conductors by a 20 μm gap. Two kind of CPWs were designed, one straight, another with a 90º corner, to apply the bias field parallel or perpendicular to the rf field, which is always directed parallel to the triangles base.

### 2.2. *Magnetization curves*

The magnetization curves shown in Fig. 2 were measured at fixed temperatures between -3000 and 3000 Oe inmagnetic field steps of 10 Oe using a QUANTUM DESIGN superconducting quantum interference device (SQUID) magnetometer. The field is applied in both directions, parallel to the base (to induce a B state) or perpendicular to it (to induce a Y state).

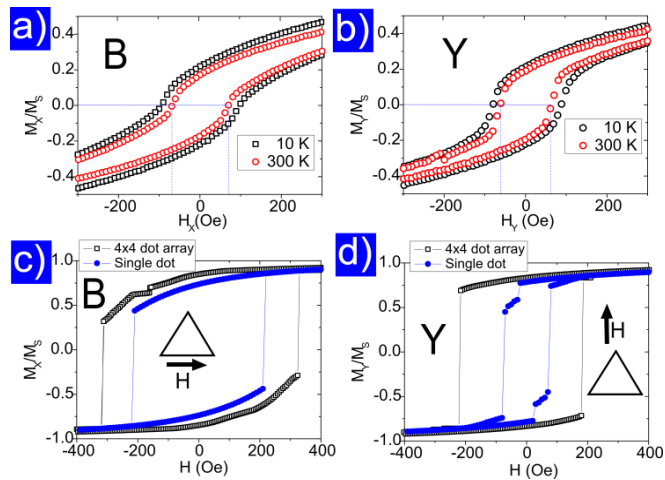

Fig. 2. Magnetization curves measured with SQUID in a) and b) and simulated in c) and d). "B" refers to buckle state with field applied parallel to the base, in a) and c). "Y" refers to Y state with field applied perpendicular to base in b) and d). Insets in a) and b) show measured data in a larger field range.

It can be noted from magnetization curves (Fig. 2) how the magnetization switching is more abrupt in the Y state (field perpendicular to the base). Also, the coercive field is lower in this case (62,5Oe in Y state, compared to 68.5 Oe in B state at T=300 K; 83,5 Oe in Y state compared to 97 Oe in B state at T=10 K).In our view, this due to shape anisotropy, and confirmed by simulations (see Fig. 2c,d), since the inclined sides of the triangle favor more strongly the magnetization alignment perpendicular rather than parallel to the base.

The measured (Fig. 2a,b) and simulated (Fig. 2c,d) hysteresis loops show some differences, mainly the "steps" present in the simulation, compared to the continuous change in measurements. It should be noted that the experimental measurementsare performed on an array with hundreds of thousands of dots (possibly with small fluctuations in shape from one dot to another). Therefore a large averaging of the switching fields of different triangles provides the smooth shape of SQUID measurements. In contrast, in the simulations, due to limitations of time and computational capabilities, both a single dot and a 4x4 array of dots are simulated. While in the measured array of dots there can be a distribution of switching fields, in the simulated array all dots switch at the same field. Also, in the measurements, the external field changes in matter of seconds, much slower than the typical relaxation time of magnetization (ns), and changes in magnetization easily and smoothly follow changes in field. On the other hand, in simulations, the external field is changed in step-like instantaneous jumps. Each of these jumps excites dynamic magnetization precession in abroad frequency range that is damped out after some nanoseconds (this is the main cause of the large computational time). These jumps, at fields close to reversal can trigger it, as opposed to the measurement case. Another source of some discrepancy is that simulations do not consider the effect of temperature.As measurements show (Fig. 2a, b)there is a decrease of total magnetization at higher temperaturesdue to thermal excitation of spin precession. This effect is not included in simulations.

In any case, it can be seen that in the Y state, at high fields both branches (reducing and increasing field) get closer, due to shape anisotropy.As will be discussed later, interdot coupling is more important in the Y state, and simulated hysteresis loops show this, being much more similar the hysteresis loops for a single dot and for a 4x4 array of dots in B state than in Y.

### 2.3. *High frequency broadband response*

The high frequency response of the dots is measured with an Agilent E8363Cvector network analyzer (VNA). First, the sample is biased with a DC field, and then a frequency sweep is performed. The triangles are excited with the magnetic field generated by the CPW when the high frequency signal of the VNApropagates through it. This rffield is parallel to the dots plane, except near the edges of the CPW, where it points out of plane.

We measure in a reflection configuration, therefore only the $S_{11}$ parameter is needed. By subtracting the measured $S_{11}$ at a reference bias field, for the rest of the fields only the changes due to magnetization in the sample are observed.

Fig. 3 shows a comparison of measurements and dynamic simulations (more details on this in the next section) at high frequencies. In figures 3 a) and 3 c) the imaginary part of differential measurements (i.e. the quantity $S_{11}(H_{i+1}) - S_{11}(H_i)$) is shown.

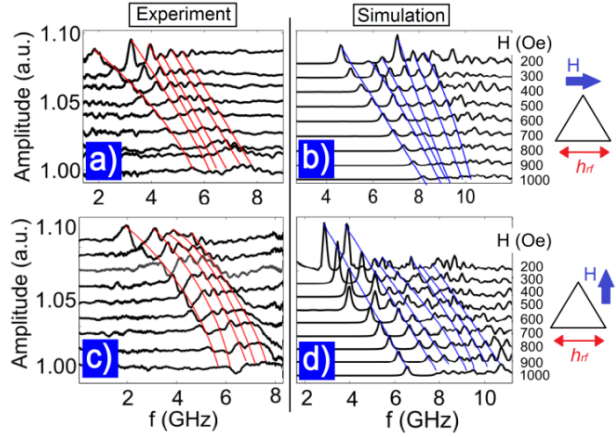

Fig. 3. a) Measured (Im{$S_{11}(H_{i+1}) - S_{11}(H_i)$}) and b) simulated (Fourier Transform of $M_X$ amplitude) broadband spectra of triangular dots in the B state. c) Measured and d) simulated broadband spectra of triangular dots in the Y state. The applied field is written besides the corresponding trace (for simulation and corresponding measurements). Traces have been separated for different fields for clarity. Color lines are guide for the eye to show the shifting of the modes.

Almost evenly spaced resonances can be observed for both field orientations (B and Y states). This resonances shift to higher frequencies when the bias field is increased. Red lines connect the same resonance position at different bias fields, and show how in the B state they are more separated at low bias than at high, whereas in the Y state their separation keeps almost constant. This seems to be in agreement with simulations.

## 3. Simulations

To understand the nature of the measured SW, micromagnetic simulations with OOMMF(Ref. 25) have been carried out. A bias field is applied homogeneously to the whole sample to reproduce the hysteresis loops, changing the value of the field step by step every 6 Oe, waiting until the system is relaxed (less than 0.1 degrees per nanosecond at every simulation cell).We used the following parameters for Py: exchange stiffness A = 1.4 ×$10^{-11}$ J/m, saturation magnetization $4\pi M_s$=10.43×$10^3$ G, Gilbert damping α = 0.01, and gyromagnetic ratio γ/2π = 2.96 MHz/Oe. The cell size used is 2.5nm x 2.5 nm in plane, and 30 nm in vertical direction.

To reproduce the dynamic response of the triangles, besides the biasing field, that is kept constant to maintain the magnetic state, a short field pulse (5 Oe amplitude, 1 psfull width half maximum) is applied. By tracking the time evolution of magnetization for long enough (15 ns in this case), with Fourier Transform the amplitude (Fig. 3b and 3d) and phase of the dynamic magnetization at each simulation cell can be obtained. With these amplitude and phase profiles, the time domain oscillation of isolated magnetization eigenmodes can be obtained.

Dispersion relation of SW (i.e. the dependence of frequency of excitations on their wave vector) can be obtained along specific directions using a two dimensional Fourier Transform to the cells comprising some path, during a certain amount of time.

### 3.1. *Static magnetic states*

Three magnetic states can exist in micron size triangular magnetic dots, namely Vortex (V) state, Buckle (B) state and Y state.

### 3.1.1. *Vortex state*

The V state appears at low fields and in plane fields easily annihilate it. Three domain walls connect the vortex core to the vertices, and can oscillate (see next section). In our 30 nm dots vortices exist only before saturating the sample, but dots thicker than 50 nm can recover them lowering the field.

### 3.1.2. *Buckle state*

An in plane magnetic field parallel to one side of a triangle orients magnetization parallel to it everywhere except the two other sides (where magnetization tries to follow the border). The transition region between close and far from these borders creates some exchange "channels" with an excess of exchange energy through which SWs can propagate.

Previous works (Refs. 12 and 13) have not identified these confined waves most probably because too small triangular dots were considered and this transition regions close to the edge spanned all over the dot.

### 3.1.3. *Y state*

If the magnetic field is applied perpendicular to one of the sides, an exchange "channel" will appear along this side, with a node separating it in two halves.

## 4. Results and Discussion

### 4.1. *Spin waves*

This section presents a comparison of the measured and the simulated spin wave spectra for different field orientations (including no applied field for the vortex state).

In general, several peaks at different resonant frequencies are observed both in experiments and simulations for every field orientation. The eigenfrequenciesshift to higher values as the absolute value of the applied field is increased. The different peaks are evenly spaced in frequencies, and they correspond to spin waves with more or less nodes along the corresponding exchange channel in each case. Some quantitative discrepancies are observed again between experiments and simulations. First of all, a relatively small shift towards higher frequencies (of about 1 GHz) has been found in the simulations with respect to the measured eigenfrequencies (Fig. 3). Other additional factors could contribute to this difference such as: (i) the saturation magnetization of the Py triangles at room temperature could be smaller than the value used in the simulations done at zero temperature, (ii) the absence of defects in the simulations, and (iii) some difference in the value of damping used in simulations relative to the real one.

#### 4.1.1. *Vortex state*

The domain walls connecting the vortex core to the triangle vertices can carry SW, referred to as Winter's magnons, already observed in other domain wall structures.[26]

The excitation ofSWalong these domain walls is in the GHz range for micron size dots, and the oscillation profiles resemble those of oscillations in a string fixed at both ends, with an increasing number of nodes as frequency increases. All three domain walls (DWs) connecting the vortex core with the vertices are equal, but, as observed in Fig.4 (top row), the DW along which magnetic moments are parallel to the rf field (in this case the vertical DW) is not excited, as expected, as only moments such that $\vec{M} \times \vec{H} \neq 0$ can suffer a torque that induces precession. Therefore, to excite oscillations in all DWs at once, a tilted rf field would be necessary.

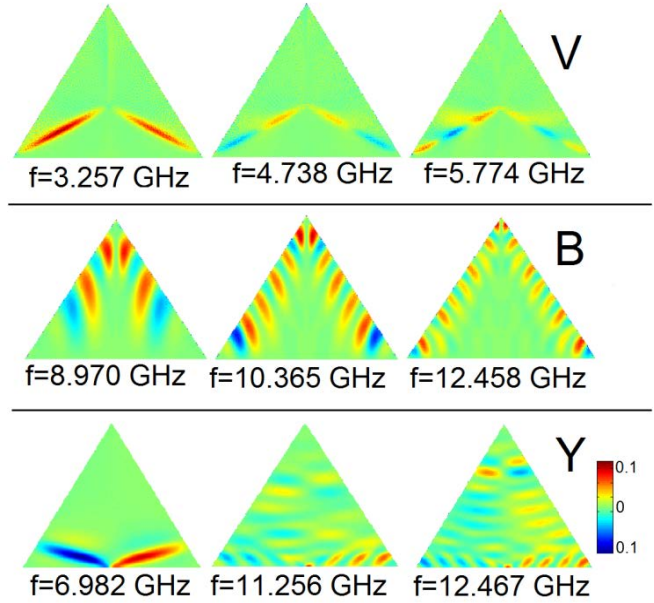

Fig. 4.Snapshot of simulated eigenmode oscillations in the V, B (H=1000 Oe) and Y (H=1000 Oe) states at different frequencies. Higher intensity of SW is observed at the exchange channels. The color scale represents $M_X/M_S$.

### 4.1.2. *Buckle state*

Quasi-one dimensional SW can propagate through the exchange channels described above (see Fig. 4, central row). Actually, when an eigenmode frequency is used to excite locally one of these dots, SW will propagate in all directions, but their amplitude will be much higher inside the channels for the highest frequency modes. These modes have the shortest wavelength, and exchange energy becomes relevant compared to long range dipolar energy.

It is to be noted that in the B state, the exchange channels are more confined close to the dot border at high bias fields, making the SW more confined to the borders, and therefore, more 1 dimensional.

Every vertex in the triangular dots presents an excess of exchange energy, and it is easy to excite SW at that point. If exchange channels are created with a bias magnetic field, these SW can propagate from one vertex to the other along them. When SWs excited in one vertex reach the other vertex, part of the incident energy is reflected, producing interferences along the channel.

If, for example, two vertices are excited with an external field at the same time, with the same amplitude and frequency, since SWs decay in amplitude as they propagate, at intermediate points of the exchange channel this interference can result in standing SW, but not near the vertices, where the amplitude of the wave propagating in one direction is larger than that of the one propagating in the opposite direction.

These SWs are detected as peaks both in the simulated and the measured spectra. Despite the shift to higher frequencies (mentioned above in the discussion of simulation results), qualitative similarities are observed: the peaks are evenly separated (an average Δf = 0.35 GHz at H=0), and the spacing between them seems to decrease at higher fields. Also, the height of the peaks lowers at higher fields, since (as already mentioned), higher fields (Fig. 3a,c) confine the exchange channels to the edges, narrowing them. Then at higher fields, SWs occupy a smaller area, and their averaged interaction with the rf field decreases, giving a lower signal. In general, the higher is the frequency, the more difficult is to observe the presence of a mode, for two main reasons.Just like increasing the applied field confines SWs to the edge, so does frequency, but in this case the exchange channel remains the same.Then, the higher frequency SWs do not spread so much all over the channel because they are more exchange energy mediated, and remain where the maximum in exchange energy density is, i.e. closer to the edges. Besides, higher frequency modes detection is more difficult because there is a larger number of maxima and minima of oscillation that average out each other. In Fig 3, middle row, it could be seen that due to asymmetry of the dot, maxima and minima of oscillation along an edge are different in size and shape. However at higher frequencies these zones are more similar and the averaged dynamic magnetization is harder to detect (the mode is more "optic" at higher frequencies, and more "acoustic" at lower).

### 4.1.3. *Y state*

SW similar to those in the B state can be obtained in the Y state, but only along the side perpendicular to the bias field. As already mentioned, to keep the symmetry in the Y state, the exchange channel at this specific side of the dot has a node where exchange energy is minimum, separating the channel in two halves. SW propagate parallel to the border, as always, but with opposite **k**. Less intense spin waves can be observed outside the exchange channel, propagating in vertical direction (see Fig. 4, bottom row). Broadband spectra reveal that the lowest frequency modes shift at certain fields, which does not occur in the B state. We attribute this shifting to the stray field of dots, that affects neighboring dots more than in the case of B state (Fig. 5). Similar to the B state, a shift to higher frequencies is present in simulations. In this case, unlike in the B state, higher frequency modes are not always harder to detect. The reason for this is that modes in this case are not confined only to the base of the triangle, but also extend to the rest of the dot (Fig 3, bottom row), and now the coupling of SWs to the rf field is not always lower, as in the B state.

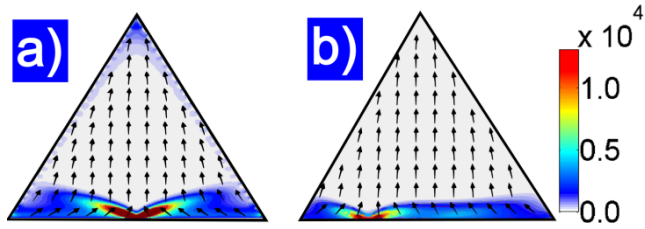

Fig.5. Exchange energy density (J/m$^3$) in colors, and magnetization distribution (arrows) for a single dot a), and for a dot coupled to others in a 4x4 array b) in the Y state.

Generally, the agreement between simulations and measurements is better in the B state than in the Y(Fig. 3).Interdot interaction is different in both states, and more important in the Y state due to shifting of the node in the exchange channel of the base, due to stray fields of neighboring dots, as shown in Fig.5.

### 4.1.4. *Dispersion relations*

As explained above, dispersion relations can be calculated from numerical simulations using 2D Fourier Transform along the desired path.
Fig. 6 shows an example for SW propagating through an exchange channel in the B state at a bias field of 1000 Oe. The parabolic shape corresponds to spin waves along the path drawn in yellow in the inset sketch. Positive **k** represents the waves propagating from the bottom left vertex to the upper vertex, and the opposite applies to negative **k**. The non symmetric dispersion curve (more intense in positive **k**) shows that waves propagate more easily towards the upper vertex than from it. This is due to the difference between the two vertices, the upper one having half of its magnetic moments pointing up and right, half of them pointing down and right, whereas the bottom left vertex has all its moments pointing up and right, and is a more robust emitter of SW.

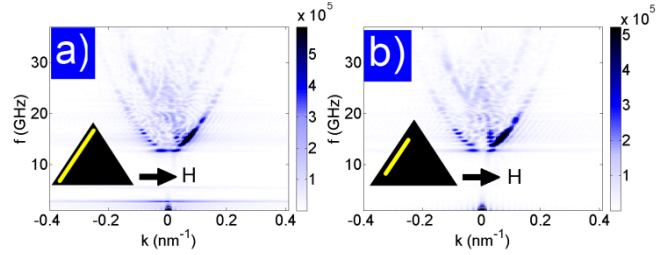

Fig. 6. Simulated dispersion relation of SW in the B state at H=1000 Oe. The inset sketch shows the path (yellow line) along which the dispersion relation is calculated. Colors represent the FFT power (a.u.). In a) the vertices are included, whereas in b) they are not. A localized low frequency mode present only in these vertices is observed at around f=2.5 GHz.

Also, when the path for calculating the dispersion relation includes the corners, a low frequency mode, independent of **k** appears. This kind of mode has already been observed in other situations (see, for example, Ref. 27) and it represents localized modes at the edges. Not including them in Fourier Transform, as in Fig. 6b, removes the presence of this mode, but not the rest.

Fig. 7 considers other cases of interest. Fig. 7a considers the case of an isosceles triangle in the B state. Everything remains qualitatively equal to the case of the equilateral shape, and SW propagate likewise through exchange channels. In Fig. 67 SW propagate in the Y state along the base exchange channels, or in vertical direction, in Fig. 7c. For both, there is not such a clear distinction between positive and negative values of **k**.

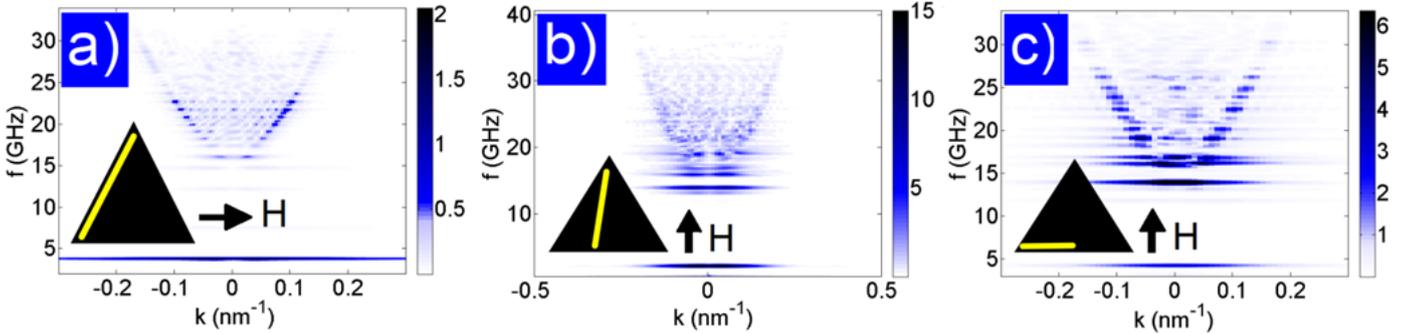

Fig. 7. Simulated dispersion relations for a) an isosceles triangle in the B state (H=1000 Oe), along an exchange channel; b) equilateral triangle in the Y state (H=1000 Oe), for vertical SW; c) equilateral triangle in the Y state (H=1000 Oe), along the exchange channel in the base (from the central node to the vertex). Colors represent the FFT power (a.u.)

## 5.  Conclusions

In conclusion, we have presented evidence for excitation and interference of quasi one dimensional spin waves propagating along the edges in 30 nm thick, 1000 nm side, triangular Py dots in the B state. These spin waves are shown to be robust with respect to variation of the dots shape from equilateral to isosceles dots with height equal to the base. Thicker dots, as observed in simulations, present a reversible vortex state that can be recovered after saturation. In this case spin waves could be excited along domain walls in the vortex state. In saturated states (B and Y) asymmetry in SW dispersion relation suggests the different SW emitting capabilities of the triangle vertices depending on the bias field direction. Interdot interaction has to be carefully considered, since it changes with the magnetic state (mainly in the Y state). Triangular dots are promising magnetic elements opening new possibilities for versatile spin wave excitation and transmission by using vertices and edge states. The observed edge spin waves could present a magnetic analogue of the edge plasmons recently predicted for graphene stripes.[28]

**Acknowledgments**

This work has been supported by the Spanish MINECO (MAT2012-32743, CONSOLIDER CSD2007-00010), Comunidad de Madrid (P2009/MAT-1726) and by the U.S. NSF, grant ECCS-0823813 grants. The authors acknowledge CCC-UAM for the computational capabilities (SVORTEX project) and K. Guslienko for discussions. A.Lara thanks Universidad Autónoma de Madrid for FPI-UAM fellowship.